**Orographic Effects on Supercell: Development and Structure, Intensity and Tracking**

Galen M. Smith[1], Yuh-Lang Lin[1,2,@], and Yevgenii Rastigejev[1,3]
[1]Department of Energy and Environmental Systems
[2]Department of Physics
[3]Department of Mathematics
North Carolina A&T State University

March 1, 2016

[@]Corresponding Author Address:  Dr. Yuh-Lang Lin, 302H Gibbs Hall, EES, North Carolina A&T State University, 1601 E. Market St., Greensboro, NC 27411.
Email: ylin@ncat.edu. Web: http://mesolab.ncat.edu

**Abstract**

Orographic effects on tornadic supercell development, propagation, and structure are investigated using Cloud Model 1 (CM1) with idealized bell-shaped mountains of various heights and a homogeneous fluid flow with a single sounding.  It is found that blocking effects are dominative compared to the terrain-induced environmental heterogeneity downwind of the mountain. The orographic effect shifted the track of the storm towards the the left of storm motion, particularly on the lee side of the mountain, when compared to the track in the case with no mountain. The terrain blocking effect also enhanced the supercells inflow, which was increased more than one hour before the storm approached the terrain peak. This allowed the central region of the storm to exhibit clouds with a greater density of hydrometeors than the control. Moreover, the enhanced inflow increased the areal extent of the supercells precipitation, which, in turn enhanced the cold pool outflow serving to enhance the storm's updraft until becoming strong enough to undercut and weaken the storm considerably. Another aspect of the orographic effects is that down slope winds produced or enhanced low-level vertical vorticity directly under the updraft when the storm approached the mountain peak.

1. **Introduction**
    Despite decades of observations and numerical simulations of supercell thunderstorms and tornados, our understanding of how orography affects tornadic supercell thunderstorms (SC) has been limited mainly to case studies of tornadic supercells that occurred over various terrain (e.g. Bluestein 2000, Homar 2003, LaPenta et al. 2005, Bosart et al. 2006, Schneider 2009). In general, the afore-mentioned case studies have found that; 1) terrain can modify winds so that there is locally enhanced storm-relative helicity, 2) upslope winds may intensify storm updrafts strengthening the mesocylone, 3) the mesocylone may be enhanced as the storm descends terrain, and 4) terrain blocking/channeling can lead to environments that are more favorable to tornadogenesis. In addition to the aforementioned environmental modifications enhancing tornadogenesis potential a few researchers have directly attributed the formation of a tornado to terrain effects (Bosart, et. al. 2006).
    Homar et. al. (2003) investigated tornadoes that occurred over complex terrain in eastern Spain on 28 AUG 1999. They used the 5$^{th}$ generation of the Pennsylvania State University - National Center for Atmospheric Research Mesoscale Model (MM5v3, Dudhia 1993, Grell et al. 1995) to

perform a triple nested simulation varying the resolution of the reference terrain and they found that during this tornadic event terrain was essential in modifying the environment to produce super cellular convection. They specifically attempted to reproduce the mesoscale environment rather than any one of the supercells or tornadoes themselves. Specifically they found that large scale terrain at the 20-50km scale enhanced super cellular convective potential, and they attributed small scale terrain features at 2-5km scales can enhance tornadogenesis potential.

Ćurić et al. (2007) investigated the effects that the terrain of the Western Morava basin, a mainly east-northeast to west-southwest oriented river valley in Serbia, had on the development of individual cumulonimbus formation in an environment with strong directional wind shear in the lowest kilometer. They investigated vertical vorticity characteristics of individual cumulonimbus clouds using the Advanced Regional Prediction System (ARPS, Xue et al. 2001), with a single domain with 600 m resolution. They concluded that vertical vorticities were enhanced at the lowest levels of the simulated clouds when terrain was included in the simulation. Additionally, they found that terrain had intensified the splitting of the simulated cloud.

Ćurić and Janc (2012) also investigated the effect of differential heating (due to complex terrain) on the evolution of hailstorm vortex pairs. This investigation incorporated both observational analysis and a numerical cloud model (Ćurić et al. 2003a, 2008). They determined that differential heading modified environmental wind shear to favor the right mover in split hail storms. This is due to stronger convection over the sunward side of terrain generating vertical vorticity that weaken the anticyclonic storm (left mover) more than the cyclonic storm (right mover). Furthermore, they find that hail point maximums occur more frequently over complex terrain.

Even though the above studies found terrain to enhance low-level rotation they were of actual tornadic supercells and the models were initialized with observations and used real terrain (although smoothed). Even though researchers do not discount that terrain influences the strength and development of supercells, the primary difficulty with observational studies is that observations are incomplete (uncertainties and/or limitations in the observation resolution lead to missing data) and that necessitates speculative attribution about the impact of terrain on the observed structure and evolution of supercells and tornados. Thus, in order to thoroughly understand the influence of terrain on supercells, an idealized model investigation is necessary.

Frame and Markowski (2006) and Reeves and Lin (2007) previously have studied the effects of mountain ridges on mesoscale convective systems (MCSs). It was found that the forward speed and depth of the outflow are affected by its passage over a terrain barrier, with the outflow slowing and thinning as the mountain crest is approached, and then accelerating and deepening rapidly in the lee of the barrier, often forming a hydraulic jump. Because the evolution of an MCS is critically tied to the behavior of the cold pool — the MCS is maintained by the continuous triggering of new cells by the cold pool — terrain-induced modifications of cold pool evolution and structure unavoidably affect the evolution and structure of the MCS. Frame and Markowski (2006) found that many MCSs weaken as they approach a mountain crest and then re-intensify in the lee of the mountain where a hydraulic jump develops in the outflow (i.e., where the outflow depth rapidly deepens).

Markowski and Dotzek (2011) have investigated the effects of idealized terrain on supercells. In their research they considered two types of idealized terrain in three-dimensional simulations. The first is a bell-shaped mountain 500 m tall and the second a gap incised into a flat topped hill 500 m tall both have a half-width of 10 km. In both situations they aimed the supercell at the location of the maximum positive vertical vorticity anomaly, as identified by an environmental simulation in which a storm was not initiated. Specifically germane to the bell-shaped mountain simulation, their results showed that the most notable change in the supercells evolution is a gradual strengthening of the mid-level and low-level updraft by upslope winds; then followed by a weakening of the updrafts but a rapid spin-up of low-level vorticity when the storm propagates over the vertical vorticity anomaly. Although Markowski and Dotzek (2011) advanced our understanding of how supercells are affected by terrain induced environmental modifications, there are several factors that were not considered in their study such as mountain height variation and its impact supercell track, intensity, structure and development.

Although these areas are not independent of one another they can be separated into a few main areas: the supercells structure and development, intensity, and track. Knowing how the terrain affects the storm helps forecasters to issue severe storm watches and warnings. Secondly, this knowledge allows forecasters to know if the storm will become more severe, change its direction of propagation, or more likely to experience tornadogenesis.

In this paper we intend to numerically investigation of the effects of an idealized bell shaped mountain on supercells. In particular, we are very interested in the scientific problems mentioned above, such as orographic effects on supercell track, intensity, and internal structures and development. Our study, on the influence of terrain on supercells, uses bell-shaped mountains of varying height that possess the same half-width. In the next section, we discuss our methodology and model selection. In section 3, we present the results of our control simulation with flat terrain; this section will describe the evolution of the environment in simulations with terrain. Section 4 describes the effects of terrain on simulated supercells, which is broken down into three subsections: first the terrain effects on the supercells track; secondly the intensity of the storm is investigated; and thirdly the structure and development of the supercell is studied. Section 5 contains our summary and closing remarks.

## 2. Methodology
### 2.1. Model Selection and Description

Our simulations utilize the Bryan Cloud model Version 1, Release 16 (CM1) (Bryan and Fritsch 2002). CM1 is a non-hydrostatic idealized numerical model designed to utilize high resolutions, particularly for severe local storms which contain deep moist convection. The governing equations that CM1 utilizes conserve mass and total energy, but they are not fully conserved in the model due to limitations in numerical integration. The CM1 introduced new equations for calculating gradients that better conserve mass and energy in simulations containing terrain and that employ stretched vertical coordinate. CM1 uses the Gal-Chen and Somerville (1975) terrain-following coordinates to map the model levels to the terrain while the model top is at constant height, and the governing equations are adapted from those described by Wicker and Skamarock (2002). The advection terms are discretized using fifth-order spatial finite difference and artificial diffusion may be applied both horizontally and vertically using separate coefficients. The sub-grid turbulence parameterization is similar to the parameterization of Deardorff (1980). CM1 has several options in microphysics parameterization schemes and the default scheme is the Morrison double- moment scheme (Morrison 2005).

### 2.2. Model Configuration and Experimental Design

The domain is $300 \times 100 \times 18$ km in the *x*, *y*, and *z* directions, respectively. In order to study the impacts of terrain a storm, the grid is stationary, instead of moving with the storm; otherwise the path of the storm will be affected. The horizontal grid spacing is 500 m; the vertical grid spacing varies from 25 m in the lowest 500 m, to constant 500 m from 11 km to 18 km (74 vertical levels total). The environments are horizontally homogeneous at the start of the simulations except in cases where storms were initialized with a warm bubble, 2 K warmer than the environment, centered 46 km north and 35 km east of the southern and western domain boundary, respectively. The warm bubble was centered at 1.4 km above the lower boundary, was 1.4 km thick and had a horizontal radius of 10 km. Simulations were run for a period of 4 h.

Simulations with storms were initialized in a way that the supercell arrived near the terrains central point at approximately the 180[th] min of simulation time (i.e. the supercell would be quasi-steady when it interacted with the terrain). In addition, the location was chosen such that the supercell propagated as close to the peak of the terrain as possible.

The terrain used in this research is centered at 200 km from the eastern boundary and 50 km from the northern and southern boundaries and is a bell-shaped mountain. The bell-shaped mountain is defined by the following:

$$h(x, y) = \frac{h_m}{\left[1+\left(\frac{x-x_0}{a}\right)^2+\left(\frac{y-y_0}{b}\right)^2\right]^{\frac{3}{2}}} \quad (1)$$

where $h_m$ is the mountain height, $a$ and $b$ are the mountain half-widths, and $(x_0, y_0)$ is the center of the mountain. The half-width is a constant 10 km in both the x and y directions and the height is varied from flat terrain to 500 m, 1000 m, and 1500 m bell-shaped mountain. Keeping the half-width the same effectively increases the terrain blocking and lifting effect.

The lower boundary is free-slip and the upper boundary utilizes a Rayleigh damping layer (Durran and Klemp 1983) in the uppermost 3 km of the model domain so that gravity waves generated by the terrain and convection are not reflected back into the domain. Lateral boundaries are open and radiative (Durran and Klemp 1983). Surface heat fluxes, atmospheric radiative heating, and the Coriolis force are set to zero for our simulations. The simulation uses the NASA-Goddard version of the Lin-Farley-Orville (LFO) microphysics parameterization scheme (Lin et al. 1983). The environments of the simulated storms are initialized with a sounding very similar to the analytic sounding of Weisman and Klemp (1982, 1984 denoted as WK82 hereafter) (Figure 1) and a warm bubble as described above. Although it has been found that models initialized with the WK82 (standard) sounding resulted in a moist absolutely unstable layer when ascending over a relatively small hill (Bryan and Fritsch 2000, Markowski and Dotzek 2011); we believe that this was due to issues with the way previous models handled momentum and energy, as our model output did not indicate a moist absolutely unstable layer generated by the terrain.

The sounding has a mixed layer convective available potential energy (MLCAPE) value of 1955 J kg$^{-1}$ and a mixed layer convective inhibition (MLCIN) of 33 J kg$^{-1}$. The environmental wind profile is defined by the analytical quarter-circle hodograph described by WK82 (Figure 1). The WK82 wind profile has a bulk shear (0–6 km shear vector magnitude) of 32 m s$^{-1}$ and storm-relative helicity (SRH) of 191 m$^2$s$^{-2}$. The *supercell composite parameter* (Thompson, et al. 2005 and 2007) of this wind profile is approximately 15. This is not surprising since the vertical moisture and wind profile for the WK82 was developed to simulate supercellular convection (although many of the included soundings in their analysis were tornadic). Moreover, the *significant tornado parameter* (STP) is greater than 2. Note that values of the STP greater than 1 are associated with the majority of tornadoes stronger than F2 while non-tornadic supercells are associated with STP values less than 1 (Thompson, et al. 2005 and 2007).

## 3. Terrain induced environmental modifications
*3.1 Environmental Simulation (Mountain Only – MTNO)*

To investigate how the environment evolved with a mountain without the presence of a storm, simulations were performed with bell-shaped mountains of 500, 1000, and 1500 m heights. One method to measure the terrain blocking effect is the moist Froude number ($F_w$) and is defined as $F_w = U/(N_w h)$ (e.g., see Lin 2007, Chen and Lin 2005, and Emanuel 1994), where $U$ is the basic wind, $N_w$ is the moist Brunt–Väisälä frequency, and $h$ is the mountain height. Both the basic wind and the Brunt–Väisälä frequency are averaged over the depth of the mountain. Changing the terrain heights effectively varied the $F_w$; 1.78, 0.89, and 0.59; for the above terrain heights respectively.

The model output for these MTNO simulations showed a general region of reduced MLCIN over and around the underlying terrain (mostly associated with a reduction in the distance from the surface to the Lifting Condensation Level (LCL)), however, as the simulation progresses, the greatest MLCIN reduction occurred just north-east of the mountain peak. Evidence of gravity waves modifying MLCAPE were present starting 90 min into the simulation as regions of alternating reductions in MLCAPE were seen radiating away from the terrain toward the east-north-east (Figure 2). Moreover, the control simulation output showed a general region of increased MLCAPE over and around the underlying terrain, with the greatest MLCAPE increase occurring near the peak of the 500 m terrain. However, as the terrain height is increased above the LCL there is an associated reduction in the MLCAPE that evolves throughout the simulation to produce lower MLCAPE over the terrain peak (Figure 2). As with the MLCAPE field, evidence of gravity waves modifying MLCIN was present with alternating regions of increased/decreased MLCIN radiating away from the terrain toward the east-north-east.

In simulations with higher mountains the modifications to MLCAPE/MLCIN were stronger, and in fact, supercellular convection was initiated by terrain induced environmental modifications (gravity waves) at approximately the $60^{th}$ and $120^{th}$ min for the 1500 and 1000 m simulations respectively. The location of the terrain induced supercell was approximately 40 km north-east of the terrain peak. It appears that this does not hinder our results as the cold pool did not significantly propagate over or around the mountain. Furthermore, the cold pool did not interact with that of the initialized storms until after our analysis is complete.

Further analysis of the low-level vorticity and wind field showed that the 1500 m terrain simulation is the only one that generated a closed pair of counter-rotating vortices (Figure 3). Although the maximum vertical vorticity generated by the 1500 m mountain was 0.013 s$^{-1}$, this was at the $60^{th}$ min of the simulation and weakened by the $180^{th}$ min. The vertical vorticity extrema, cyclonic and anticyclonic, is ±0.001 s$^{-1}$ for the 500 m and ±0.0025 s$^{-1}$ for the 1000 m mountain simulations.

*3.2 No Mountain Control Simulation (NMTN)*

To establish a baseline of how CM1 simulates a supercell thunderstorm we performed a simulation without terrain. This simulation was initiated by the same warm bubble as those for the MTN (mountain) cases. This will also allow us to better isolate the effects of terrain. The control simulation produced well-defined right-moving supercell thunderstorms with sustained midlevel, 5 km Above Ground Level (AGL), updrafts (downdrafts) exceeding 30 m s$^{-1}$ (15 m s$^{-1}$) and low-level, 500 m AGL, updrafts (downdrafts) exceeding 5 m s$^{-1}$ (10 m s$^{-1}$). Organization of midlevel rotation (vertical vorticity 0.003 s$^{-1}$) was incipient within 30 min of simulation time and was well organized by 45 min (with vertical vorticity 0.02 s$^{-1}$), see Figure 4. Midlevel cyclones were sustained throughout the end of the simulations and cyclic intensity is seen as indicated by the 1 km AGL updraft strength, see Figure 4; consistent with observations (Burgess et al. 1982; Beck et al. 2006) and previous numerical simulations (Klemp and Rotunno 1983; Wicker and Wilhelmson 1995). Simulated radar reflectivity gives a clear indication of the classic supercell structure at the $105^{th}$ min (Fig. 5). Figure 5 also shows that the midlevel rotation is aligned with the updraft, indicated by rotating winds aligned with the bounded weak echo region (BWER), which makes the storm more conducive to tornadogenesis. The control simulation vertical vorticities ranged from 0.02-0.05 $s^{-1}$ 50 m AGL, from the end of the $60^{th}$ and out to the $180^{th}$ min. This storm propagates eastward at approximately 15 m $s^{-1}$, with small north to south variation in the location of the storm staying within approximately ±5 km north-south from the location of the warm bubble used to initiate convection. The NMTN also had an anticyclonic left-moving storm that propagated out of the domain by the $120^{th}$ min.

**4. Mountain Simulations (MTN)**

In the following we focus on the investigation of orographic effects on supercell thunderstorm; structure and development, intensity, and track.

*4.1 Orographic Effects on Supercell Structure and Development*

Overall the initial development of these supercells are quite similar for the NMTN and MTN simulations, the storms undergo more or less identical storm development during the first 60 min of the simulation, and begin to exhibit the structure of the classical High Precipitation Supercell conceptual model (Lemon and Doswell 1979). They remain structurally quite similar throughout the maturing phase, ~90 min, although there is increased rainfall area in the 1000 and 1500 m MTN cases, (M1000 and M1500 respectively). The structure of these storms diverged significantly by the $150^{th}$ min.

A remarkable difference between the MTN cases and the NMTN case at the $150^{th}$ min is the distribution of hydrometeors within the cloud. NMTN case has a distribution of cloud water and ice water that is approximately twice as large in horizontal extent as compared to the MTN simulations (Figure 6). The cloud size is partially attributable to the midlevel winds advecting the cloud and ice hydrometeors towards the east as the eastward winds are stronger in the NMTN case than in the MTN simulations, see Figure 6 and Figure **7**. Furthermore, the cloud hydrometeor differences are also noticeable, which are related to the increased rainfall in MTN simulations reducing the overall amount of water available. The cloud that is indicated by reflectivity in the MTN simulations is accounted for as a mixture of snow and graupel hydrometeors. The decreased cloud region at the lower levels is also attributable to the updraft core being larger, stronger, and better organized in the MTN cases than that of the NMTN case. The stronger updraft produced a larger over shooting top and allowed the anvil cloud to become deeper on the upwind side of the storm.

The hydrometeor densities in and around the main updraft region of MTN cases is higher than that of the NMTN case (not shown). The distribution of hydrometeors is primarily affected by the redirection of additional air into the storm modifying the storms structure; consistent with the findings of Curic and Janc (2012) in which they found that differential heating associated with terrain could favor an enhanced right moving storm and intern alter the hydrometeor distribution.

The augmented air-flow into the supercell produced rain over a greater areal extent and a more continuous rainfall in the MTN cases. The rainfall area is also shifted towards the north in relation to the NMTN case. This is consistent with our findings that the track was shifted towards the north in MTN cases. The increased areal extent of rain allowed the cold pool to strengthen and intensify the storm until the gust front undercut the updraft which weakened the storms midlevel updraft considerably (Figure 6d and Figure **7**d). The M1500 storm reorganized once it propagated away from the area where the gust front undercut the supercell.

The NMTN simulated storm developed a low pressure in the 6 to 10 km layer immediately east of the main updraft (Figure 6a), indicated by divergent winds associated with precipitation loading. As this low pressure strengthened (Figure 7a) winds from the main updraft were turned toward the east, until the main updraft was effectively split horizontally at approximately 8 km (Figure 8a). This shifted the cloud base to the east of the main updraft and reduced the rain rate as indicated by the reduction of reflectivity (Figure 6a-9a). Interestingly, the upper-level updraft intensified while the mid-level updraft weakened (Figure 8a and 10a).

The M500 simulated storm maintained its updraft size and strength more than those of the other simulations (Figure 6-9). The updraft became larger and stronger as the storm approached the mountain peak and the gust front converged with the winds which were diverted by the mountain. However, the larger updraft started to ingest air from its cold pool (Figure 6b and 10b) essentially offsetting the enhancement of the upslope winds coupled with the main updraft. This effectively produced a storm that varied less structurally throughout the time the storm was interacting with the terrain.

The M1000 simulated updraft intensified slightly and became more upright and the upwind part of the anvil cloud shallows as it approached the mountain (Figure 6c and 8c). However, as the storm propagated up the mountain the blocking effects on both the storm inflow and the cold pool weakened its updraft considerably (compare the low-level in-flow in Figure 7c and 9c). The blocking effect also reduced the storm's propagation speed which allowed the rear flank downdraft to interact with the storm's updraft, further weakening the storms main updraft (Figure 8c). Once the storm propagated to the lee side of the mountain, downslope winds coupled with the storms cold pool to enhance lifting of the lee side convergence region and the storms updraft became much larger (Figure 9c). The mid-level structure of the storm at the 195$^{th}$ min resembled that of the M500 simulation (Figure 9b and c). Once on the lee side of the mountain, the storm started to ingest cool dense air, which was associated with storms triggered further to the west of the mountain, and dissipated quickly after this time (not shown).

The propagation of the M1500 simulated storm was slowed when it started to interact with the terrain and thus it did not propagate past the mountain during the same time interval as that of the storms simulated in the NMTN or other mountain cases. The storm of M1500 was also slowed due to the reduction in the strength of the updraft associated with its ingestion of air from its cold pool and weakened considerably (Figure 6d and 9d). The ingesting of the cooler air from the cold pool also reduced the amount of precipitation, weakened the cold pool and allowed the storm to propagate out ahead of the gust front and re-intensified quite rapidly (Figure 8d). Once the supercell propagated further behind the mountain the inflow was blocked at low levels and the inflow jet was essentially cut off from the storm (Figure 9d). Similar to the M1000 simulation the mid-level updraft widened considerably at 195$^{th}$ min. Once the storm propagated over the lee side convergence zone, the inflow becomes unblocked and the storm started to re-intensify up until the point that it started to ingest cold air from the terrain initiated storms, as mentioned in section 3.1 (not shown).

4.2 *Orographic Effects on Supercell Intensity*

Our intensity investigation will primarily focus on the strength of the updraft (downdraft) and the vorticity in the mesocyclone at mid-levels (5 km AGL) and low levels (500 m AGL); The impacts on tangential wind speed, how well the vortex is formed, and the strength of the gust front will also be discussed.

The supercell in all simulations exhibited cyclic intensification and decay, consistent with observations (Burgess et al. 1982, Beck et al. 2006) and previous numerical simulations (Klemp and Rotunno 1983, Wicker and Wilhelmson 1995). Although all simulations exhibited cyclic intensity the timing of the peak intensities was altered by the terrain such that it was shortened from approximately 75 min in the control case (NMTN) to 60 min in case with terrain (MTN). In both cases, there were three intensity peaks produced throughout the simulation. The change in the intensity cycle appears due to an increase in the storms inflow rather than the storms updraft coupling with the upslope winds of the terrain, as the second cycle peak is simulated before the upslope wind could become significant. After the second intensity peak of the NMTN and M500 simulations surface vorticity weaken whereas the M1000 and M1500 simulations intensify as the storms couple with upslope winds. On the lee side of the mountain the storms low-level updraft of MTN simulations weakens most notably in the M1500 simulation, while the M1000 and M1500 weaken considerably after the $210^{th}$ min as they encounter the area of reduced CAPE associated with the outflow of storms triggered on the lee side of the terrain. Although the first two intensity peaks in the MTN storms are stronger than that of NMTN storm the MTN storms are weaker at the third intensity peak.

Although a complete account of all minor variances of the terrain simulations from the NMTN simulation would be exhaustive and tedious, as they start after the $60^{th}$ min of simulation, very early from when the terrain effects become significant (nearly 100 km from the peak of the mountain). We look and the differences in the near surface vertical vorticities, the low and mid-level updraft, and the speed of the gust front (a measure of the cold pool intensity) for the time interval from the $165^{th}$ to the $210^{th}$ min.

At the $165^{th}$ min, the 1 km AGL updraft of the NTMN simulation was 15 m s$^{-1}$ while the M500, M1000, and M1500 simulations were 9, 12, and 15 m s$^{-1}$, respectively (Table 1). The decrease in updraft velocity in the M500 and M1000 is most likely attributable to the hydrometeor density being higher in the main updraft resulting in precipitation loading. The downdraft for the M500 and M1000 simulations is stronger than the NMTN or M1500 simulations, 15 vs. 9 m s$^{-1}$ respectively, as with the updrafts being lower the downdrafts are stronger due to the precipitation loading effect. At 5 km AGL the strongest updraft of the control simulation was 35 m s$^{-1}$, and is 30 m s$^{-1}$ for all three terrain simulations. The downdrafts at this altitude are 15, 10, 15, and 5 m s$^{-1}$ for the NMTN, M500, M1000, and M1500 cases, respectively. The speed of the gust front in the NMTN and M1000 simulations is 33 m s$^{-1}$, while the M500 and M1500 storms are 35 and 45 m s$^{-1}$, respectively.

At the $180^{th}$ min the M500 and NMTN simulations surface vorticity weakened by 0.011 and 0.004 s$^{-1}$, respectively. There was little change in the M1000 simulation vorticity and the M1500 simulation surface vorticity strengthened to 0.053 s$^{-1}$ (Table 1). This increase in vorticity for the M1500 simulation was not due to the supercell coupling with the terrain induced vortex generated on the lee side of the mountain, as the storm's location is still relatively far from the location of the lee side vortex, ~30 km. The vorticity enhancement is due to stretching and terrain blocking effects physically redirecting air flow. The enhancement due to vorticity stretching is evident as the 1 km AGL updraft strength increases from 15 to 18 m s$^{-1}$ during this time. Interestingly, the updraft of the M500 simulation increased in strength from 9 to 12 m s$^{-1}$, however in this simulation the surface vorticity decreased (Table 1), due to weaker coupling of the storm's updraft with the upslope winds and reduced blocking effect not channeling the winds such that the vertical vorticity would be enhanced. At upper levels the updraft has strengthened for the M1000 and M1500 simulations by 10 and 5 m s$^{-1}$ respectively and actually decreased for the M500 simulation. There was no change in the speed of the gust front for the NMTN or M1000 simulations; the gust fronts simulated in M500 and M1500 simulations were weakened by approximately 10 and 5 m s$^{-1}$, respectively.

The increase or decrease in the updraft is attributable to two effects, the first and strongest contributor was the terrain blocking effect, which channeled air into the storm and coupled of the updrafts with the upslope winds. The turning of the winds increases the inflow wind speed from ~10 m s$^{-1}$ for the NMNT storm to 16, 14, 17 m s$^{-1}$ for the M500, M1000, and M1500 storms respectively, just as the storm is encountering the terrain. These together increase the precipitation rate (the rain coverage is increased in the M1000 and M1500 m simulations compared to that of NMTN and M500 simulations) (not shown) and strengthen the cold pool and intern the strength of the gust front. As expected higher terrain heights allowed the storms to generate consistently more rain.

The low level vorticity is strongest and most organized in the M500 and M1500 simulations at the 180 min; the updraft is also aligned with the vertical vorticity (Figure 10). In addition to the vertical vorticity aligning with the updraft the down slope winds enhance the vertical vorticity by accelerating the horizontal winds on the northern section of the storm's updraft. It is possible that the M1000 simulation also experienced this enhancement of vertical vorticity by the downslope winds on the northern section of the storm's updraft; however, the storms Rear Flank Downdraft proximity weakens the low level updraft considerably (Figure 10).

*4.3 Investigation of Methods for Tracking Supercell Thunderstorms and Orographic Effects*

A selection of parameters for determining the accurate location of a supercell thunderstorm is necessary for determining the track. First—assuming a supercell is present—we try to identify the location of the maximum updraft velocity which would yield a good track representative of the supercell's location. Although this provided a good starting point, the track was rather rough during the early part of the simulations when one would expect the simulations to be nearly identical (Figure 11 a). Next, we identify the track using the classic identifier of a supercell the rotating updraft; this parameter is the updraft velocity, at 500 and 1000 m AGL, multiplied by the vertical vorticity at that level. This provided a smoother track than that identified by the maximum updraft alone. Using the classic supercell identifier we can conclude that the track is shifted towards the north in simulations with increasing mountain height.

We continue our investigation along these lines, and use the location of maximum updraft helicity (UH) to determine the supercells location, UH is a parameter that has recently been used to identify the areas where convective storms are more likely to occur (Kain et al. 2008). UH has proved useful in its ability to detect areas more likely to exhibit convection in model output (Sobash et al. 2008). Although the UH did indicate that a storm was in the approximate vicinity of the supercell, the identified track was rather sporadic and produced quite an erratic track, this ruled out the usage UH alone as a supercell tracking method. The storm track was initially smooth during the storm's development phase; however the track became erratic throughout the rest of the simulation (not shown).

The next parameter used to identify the supercell's track was the maximum UH multiplied by the updraft velocity (UHW), which makes the track smoother than that identified by UH. The storm location was identified by tracing the maximum UHW. Again, we used the updraft velocities at 500 and 1000 m AGL. UHW noticeably improved the track and we can say that as the Froude Number decreases the supercell track is shifted towards the north, particularly at lower levels. The combination of the updraft with the updraft helicity produced a smooth track that was free of significant jumps and was more consistent at the two heights used to identify the supercells track.

Of the parameters used to identify the track of a supercell our UHW parameter yielded the best track; both based on smoothness and consistency (between different levels). Following closely after the UHW, the updraft strength produced the smoothest track as long as supercells are known to exist. The classic definition of a rotating updraft produced good results in track identification it produced jumps that were uncharacteristic of storm propagation. Interestingly the updraft helicity parameter yielded the poorest track identification with erratic track identification just after the initial strengthening phase.

## 5. Concluding Remarks

The effects of idealized, bell-shaped mountains on supercell thunderstorms were investigated in this study. The mountains produced gravity waves that modified the downwind environment by producing alternating reductions and increases in the amounts of moisture, MLCAPE, and MLCIN. The simulations with higher mountains, such as mountain heights of 1000 m (M1000) and 1500 m (M1500), produced gravity waves that had enough vertical motion to initiate convection near the $120^{th}$ and $60^{th}$ min respectively. Cold outflow from these storms reached the lee side at approximately the $225^{th}$ and $180^{th}$ min for the M1000 and M1500 simulations, respectively. Although these storms produced large environmental modifications our analysis was focused before these effects could influence the investigated supercells.

Several combinations of variables were used to create parameters for the identification of a supercell's location. Although the updraft helicity (UH) indicated the general vicinity of the supercell the identified track that was rather erratic. Other parameters that were used based on the characteristics of supercells yeilded smoother tracks; however the maximum UHW (UH multiplied by the updraft velocity) produced the smoothest tracks and tracks that were the most similar far from the mountain where the terrain effects are minimal. Using the maximum UHW we identified that increasing the mountain height shifted the tracks of supercells towards the north.

The intensity of supercells was cyclic in all simulations; however the period between intensity peaks was reduced in mountain (MTN) cases as compared to the no mountain (NMTN) case. The intensity, structure and development of the storms were mainly a result of the mountain directing an increased amount of environmental air into the storms inflow. This created differences in the distributions of hydrometeors and increased the rainfall areal extent. This allowed the cold pool to be stronger in the MTN simulations, most notably when the cold pool undercut the M1500 storm.

Airflow was also modified such that vorticity was generated and/or intensified when approaching the mountain peak. The near surface rotation of the M500 storm intensified as it approached the mountain peak. The M1000 storms propagation speed was reduced as it crossed the terrain, which allowed the storm's rear flank downdraft to run into the storms low-level updraft and reduced the near surface vorticity greatly. The M1500 storm experienced a greater reduction storm motion, however, its' rear flank downdraft was farther away from its updraft and its intensity was not affected in the same manner as M1000. The M1500 storm propagated around to the north of the mountain peak and its' cold pool worked in conjunction with the terrain to block the storm's inflow and causing the storm to weaken considerably until it propagated into the lee side convergence region.

Although these simulations did not produce tornadic supercells as the grid spacing was too coarse to reproduce such systems, model output noted (in CM1's log files) several instances throughout the MTN simulations where vertical vorticity was greater than 0.1 $s^{-1}$ at the lowest model level (12.5 m). We believe that tornadogenesis could occur if the simulations were run at higher resolutions. We have shown that blocking effects may direct additional air into the storms inflow and enhance low-level vorticity along the gust front and that these blocking effects are far more important that the than the environmental modifications, especially since we observed these differences before the storm even interacted with the environmental modifications on the lee side of the mountain. The direction of additional moist air into the storm is particularly of interest to now/forecasting because this increases the precipitation amount and was observed far from the mountain and could increase the likelihood of flash flooding. The M1000 and M1500 simulations initiated supercellular convection that reduced the MLCAPE and increased the MLCIN far more than the gravity waves excited by the mountain and indeed when the simulated storms propagated into this region they quickly dissipated.

Areas where this study could be extended in the future are to vary the arrival time of the storm to investigate the terrain effects on developing or mature storms. Additionally, the track and/or terrain configuration could be modified to test the robustness of our conclusion that the terrain blocking effects are more important that the environmental modifications. This area is still very much unexplored and additional research is needed.

**Acknowledgments**
We would like to thank the National Oceanic and Atmospheric Administration (NOAA) Educational Partnership Program and the NOAA Earth System Research Laboratory for the use of their ZEUS supercomputer to conduct these simulations. Dr. George Bryan and NCAR are appreciated for allowing us to use CM1 and NCL, respectively. We thank IGES for the use of their Grid Analysis and Display System (GrADS) plotting software. This research was partially supported by the NOAA Cooperative Agreement No: NA06OAR4810187, and National Science Foundation Awards AGS-1265783 and HRD-1036563.

**Table 1:** Selected Variables for intensity comparison from 165 - 210 min

| Simulated Time | Case  | w1km (m s$^{-1}$) | w5km (m s$^{-1}$) | Gust Front (m s$^{-1}$) | Max θ' (K) | Surface Vertical Vorticity (s$^{-1}$) |
|---|---|---|---|---|---|---|
| 165 min | NMTN  | 15/-9  | 35/-15 | 33 | -8  | 0.038 |
|         | M500  | 9/-15  | 35/-15 | 40 | -7  | 0.031 |
|         | M1000 | 12/-15 | 30/-15 | 33 | -8  | 0.025 |
|         | M1500 | 15/-9  | 30/-5  | 45 | -8  | 0.038 |
| 180 min | NMTN  | 12/-15 | 35/-10 | 33 | -11 | 0.034 |
|         | M500  | 12/-12 | 30/-20 | 30 | -9  | 0.042 |
|         | M1000 | 12/-12 | 40/-10 | 33 | -8  | 0.027 |
|         | M1500 | 18/-9  | 35/-15 | 40 | -8  | 0.053 |
| 195 min | NMTN  | 9/-12  | 30/-15 | 30 | -8  | 0.023 |
|         | M500  | 9/-15  | 30/-10 | 33 | -8  | 0.023 |
|         | M1000 | 10/-10 | 30/-15 | 27 | -8  | 0.031 |
|         | M1500 | 12/-10 | 30/-10 | 35 | -8  | 0.035 |
| 210 min | NMTN  | 10/-10 | 30/-15 | 27 | -8  | 0.027 |
|         | M500  | 10/-12 | 30/-10 | 33 | -10 | 0.018 |
|         | M1000 | 10/-14 | 35/-15 | 24 | -7  | 0.02  |
|         | M1500 | 8/-8   | 30/-10 | 30 | -8  | 0.036 |

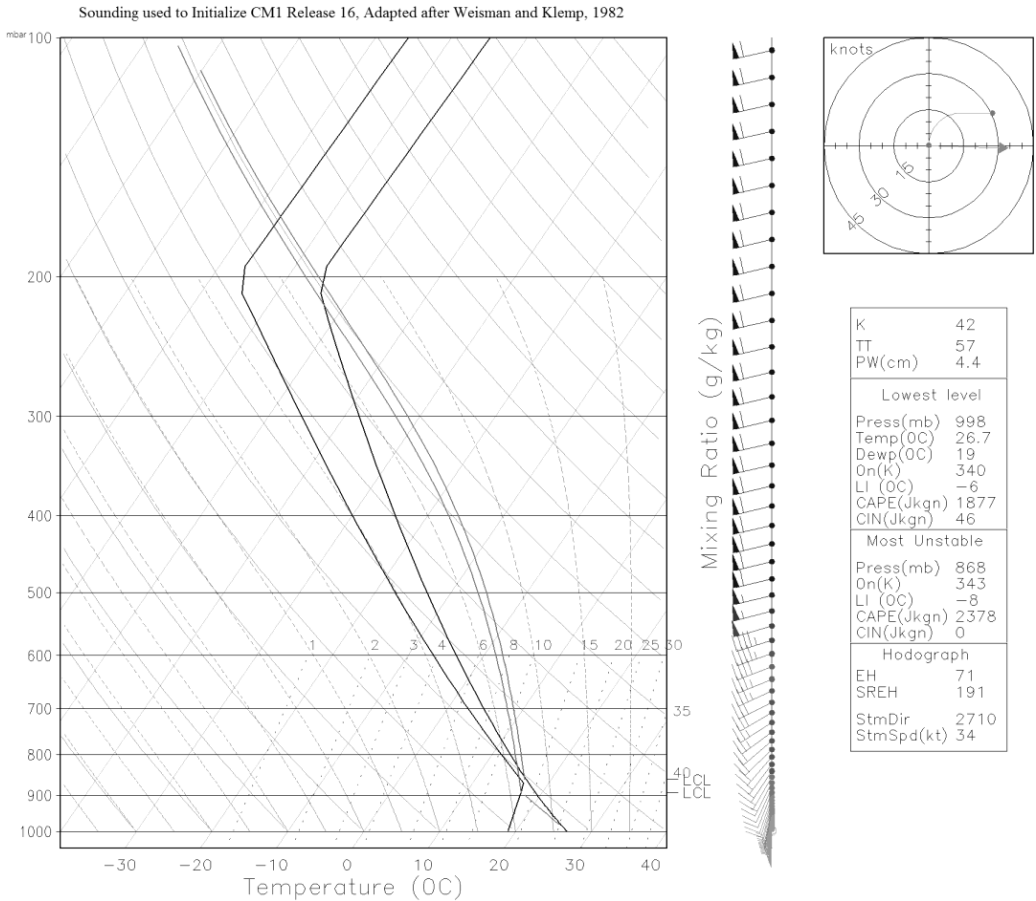

Figure 1: The sounding and wind profile used to initialize simulations in this study (Adapted after Weisman and Klemp, 1982). The hodograph can be seen in the upper right corner and several indices are indicated to the right of the wind profile. The black lines represent the dew point temperature and the temperature, left and right respectively. The grey lines represent the surface parcel ascent for the lowest level and the most unstable level, left and right respectively.

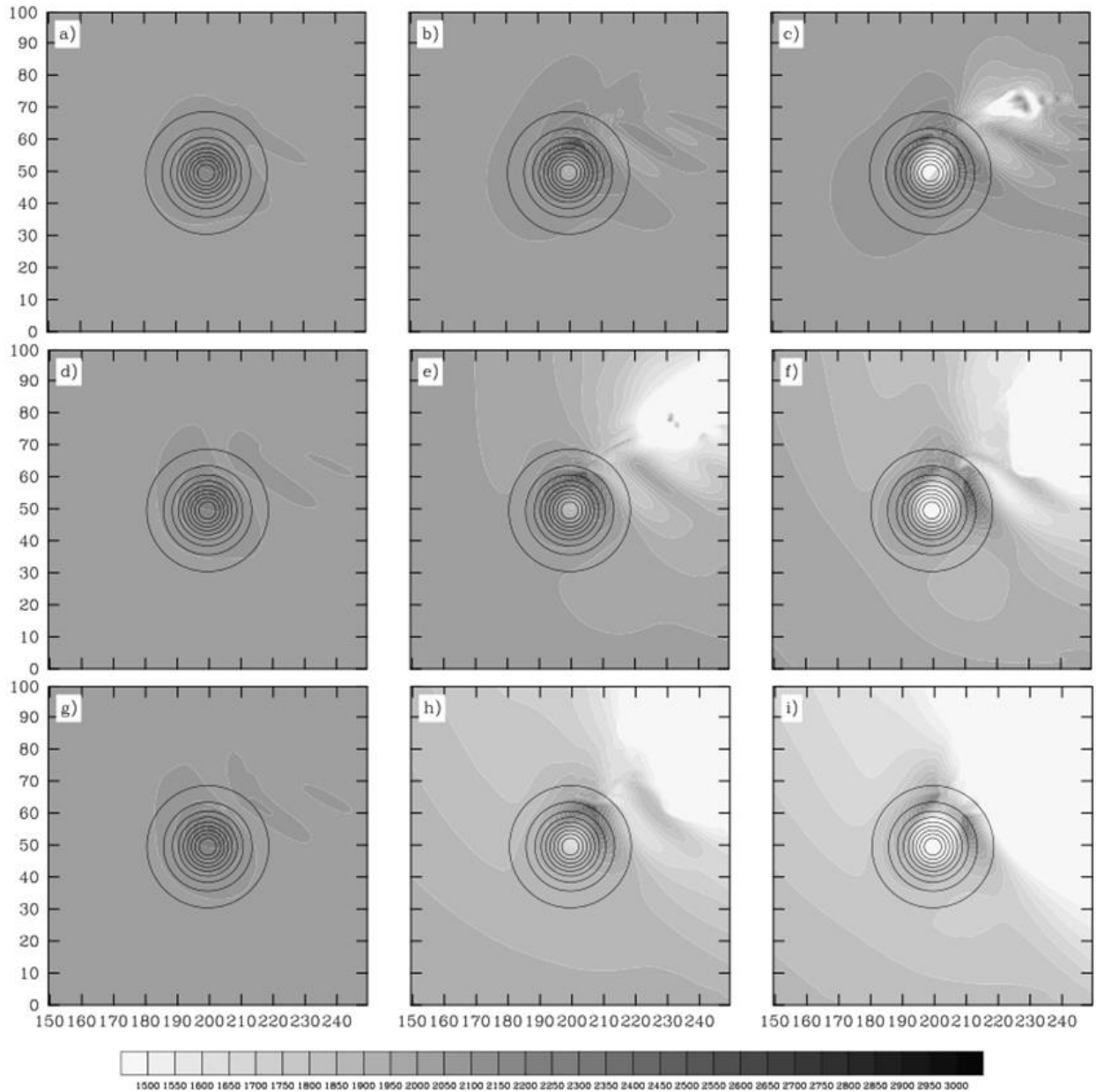

Figure 2: MLCAPE for simulations with the columns panels represent $h_m$ = 500 m, 1000 m, and 1500 m (left to right); and the row panels represent different times at 60, 120, and 180 mins. Note the region of depleted MLCAPE in c, e, f, h, and i are associated with storms triggered by terrain induced gravity waves. The contours represent the percent reduction in height; each contour from the peak represents a 10% reduction.

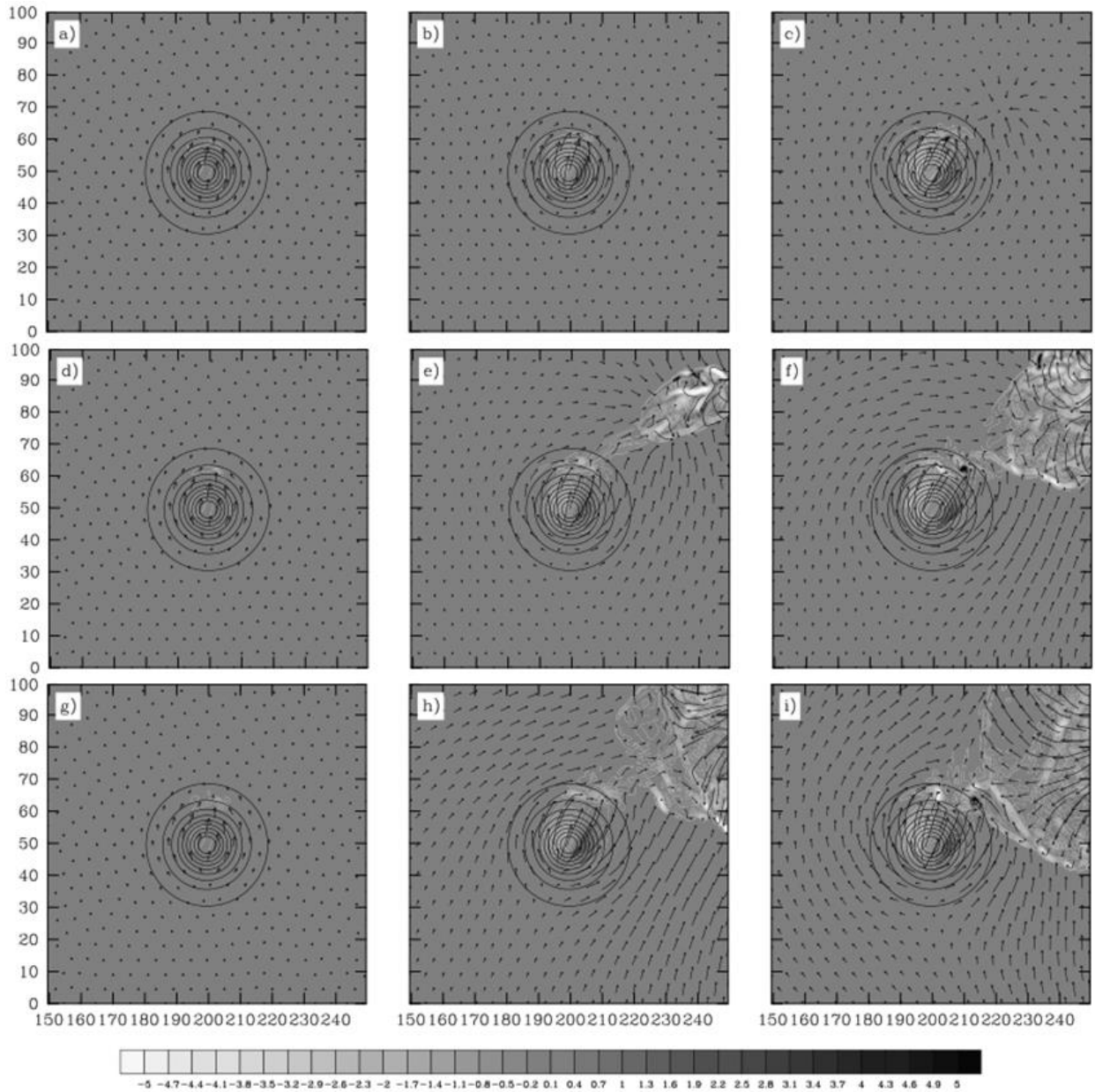

Figure 3: Low-level vorticity and horizontal wind vectors for simulations with column panels represent $h_m$ = 500, 1000, and 1500 m (left to right); and the row panels represent different times 60, 120, and 180 mins. Note the region of convergence associated with the outflow from the storms initiated to the north-east of the terrain. The 1500 m mountain was the only one that generated a closed pair of counter rotating vortices with vertical vorticities of 0.008 and -0.006 $s^{-1}$, respectively at the 180 min.

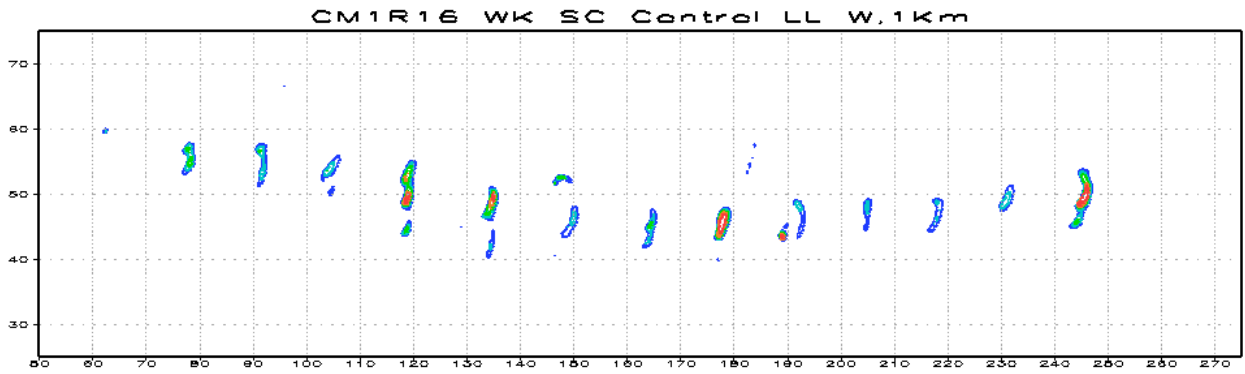

Figure 4: The cyclic nature of the simulated supercell thunderstorm can be seen in the strengthening and weakening of the 1km AGL updraft. Contours are blue, light blue, green, orange, and red representing the 10, 12.5, 15, 17.5, and 20 m s$^{-1}$ values respectively.

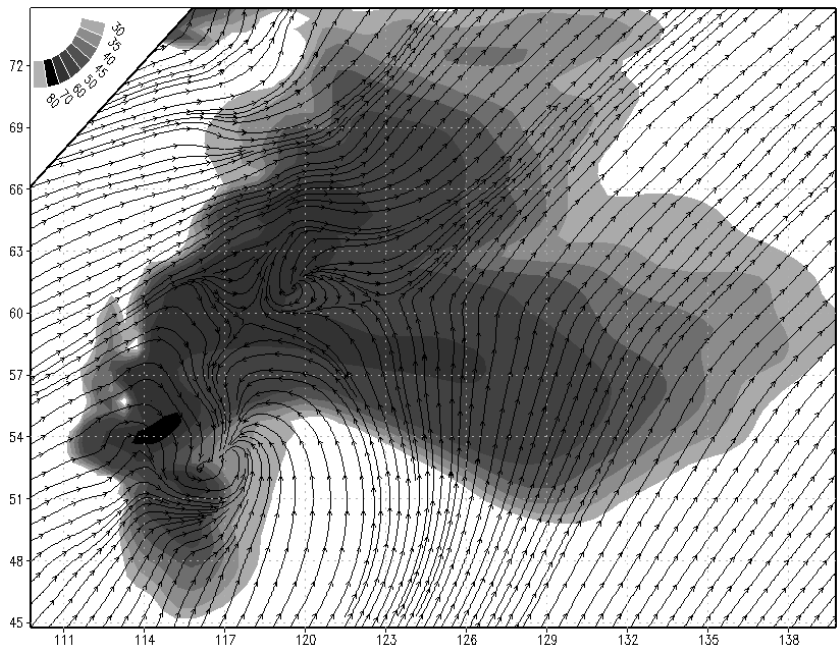

Figure 5: Reflectivity and wind stream-lines for the no-terrain control simulation (NMTN) at the 105 min. Note that the mid-level rotation is aligned with the BWER.

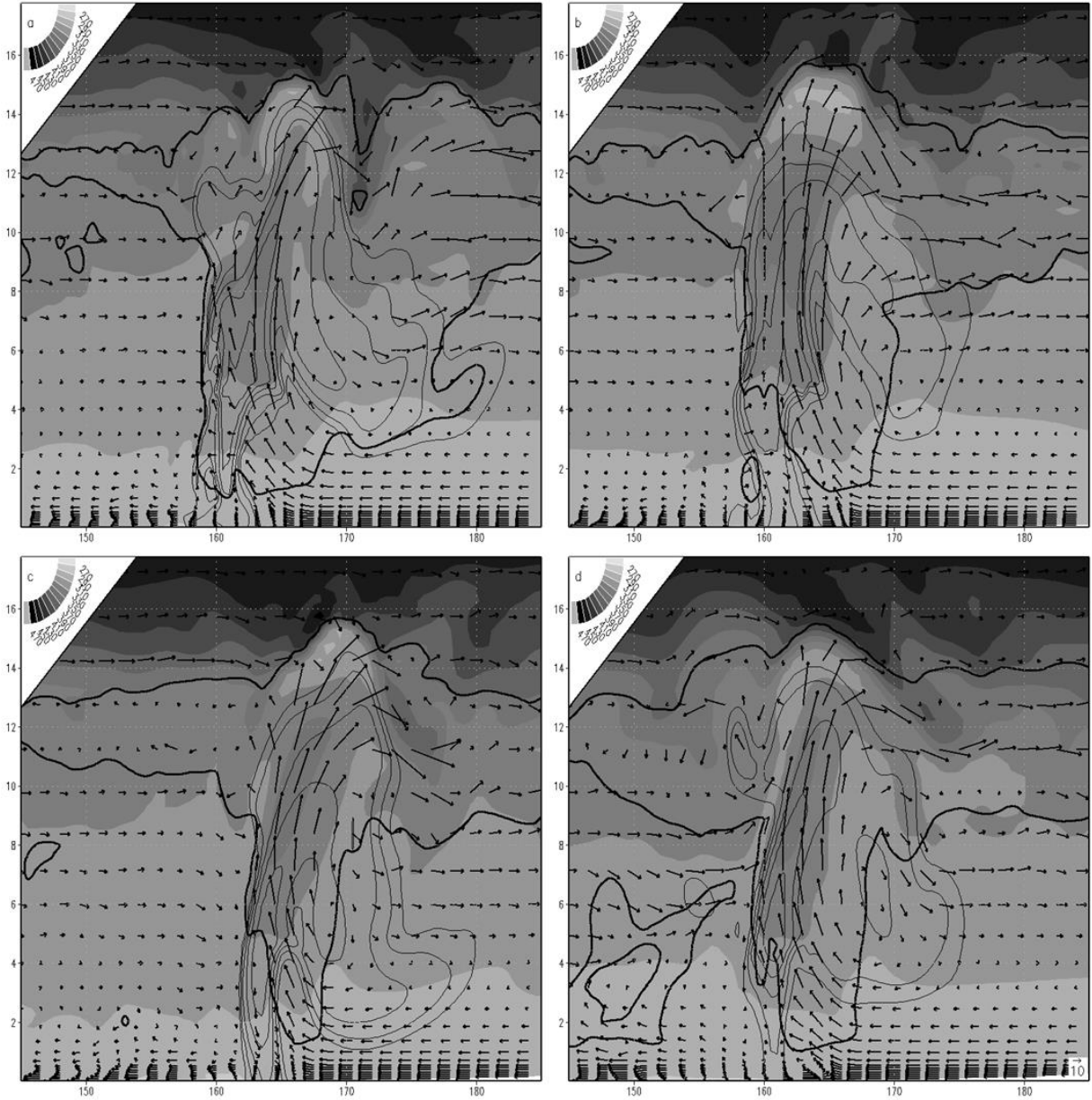

Figure 6: Zonal cross section of theta, reflectivity, cloud outline, and wind vectors, at the 150 min in cases NMTN, M500, M1000, and M1500 simulations a, b, c, and d respectively. Theta is shaded. Reflectivity values start at 50 dBZ and are contoured every 5 dBZ (thin contours). The cloud boundary is indicated by the 0.5 g kg$^{-1}$ cloud water/ ice mixing ratio. The reference vector is in the lower right corner of panel d and is the same for all panels. Cross section is along the direction of propagation (east-west) and is at the point of maximum UHW.

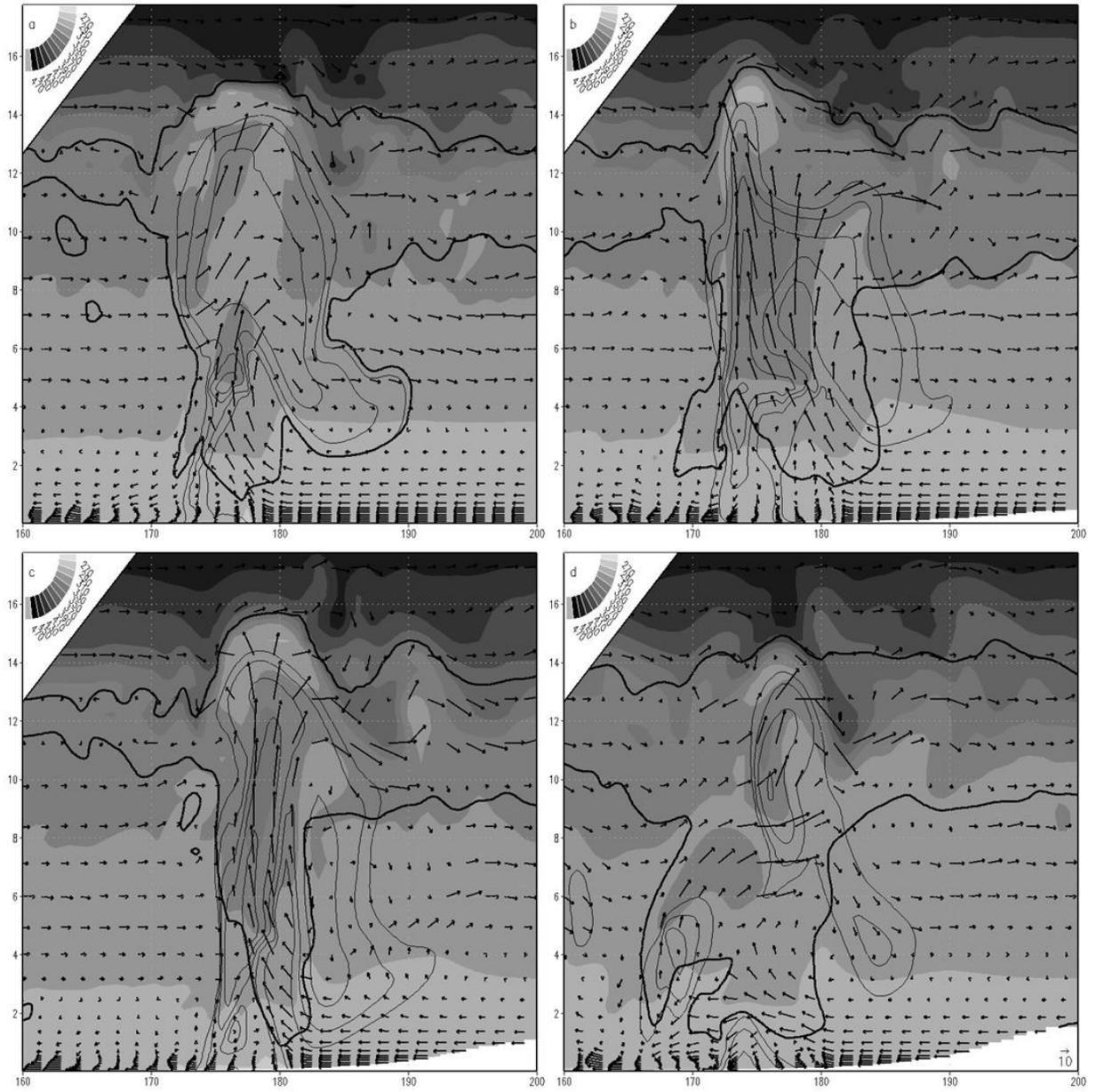

Figure 7: As in Figure 6 except for 165 min.

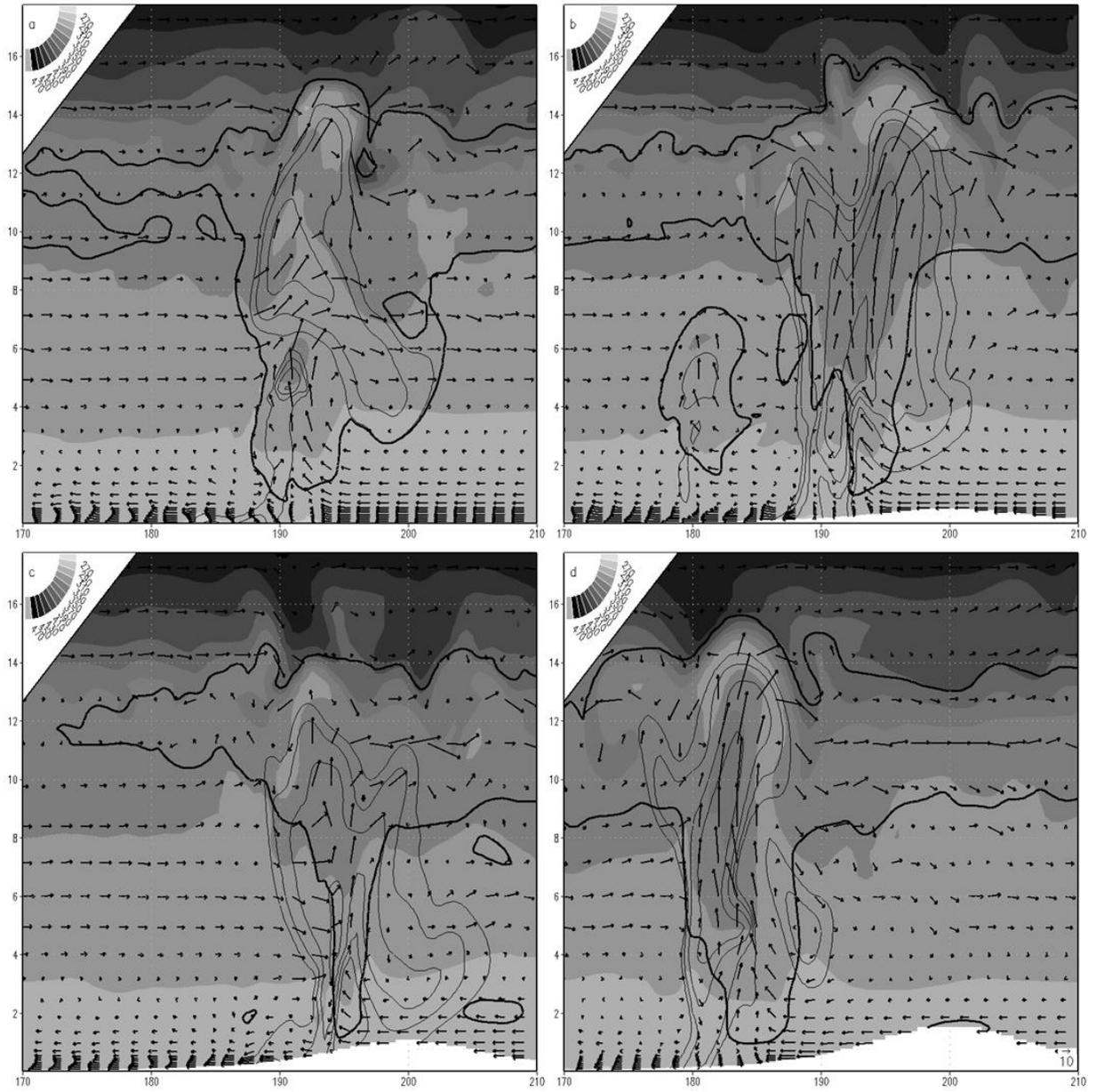

Figure 8: As in Figure 6 except for 180 min.

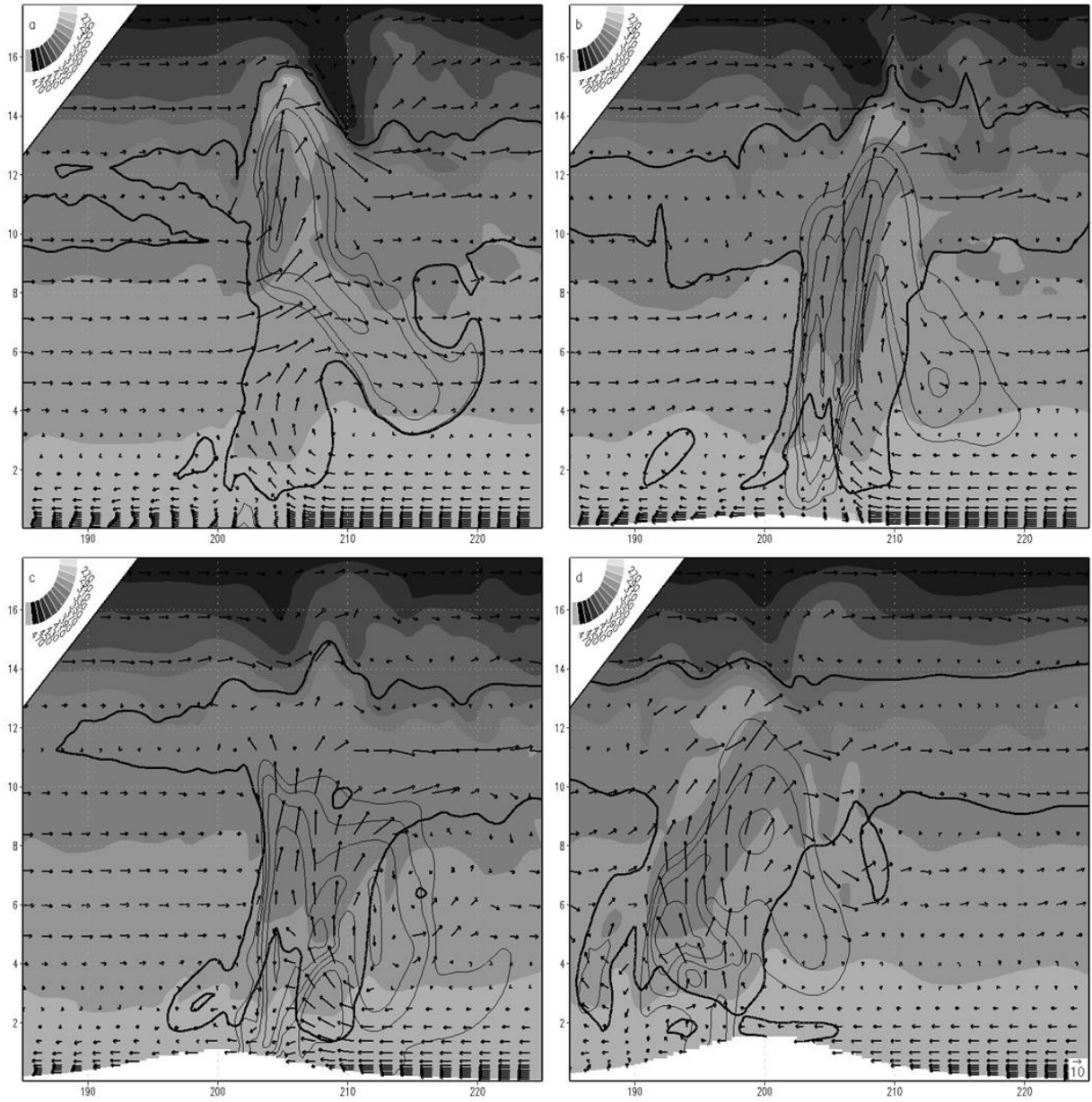

Figure 9: As in Figure 6 except for the 195 min.

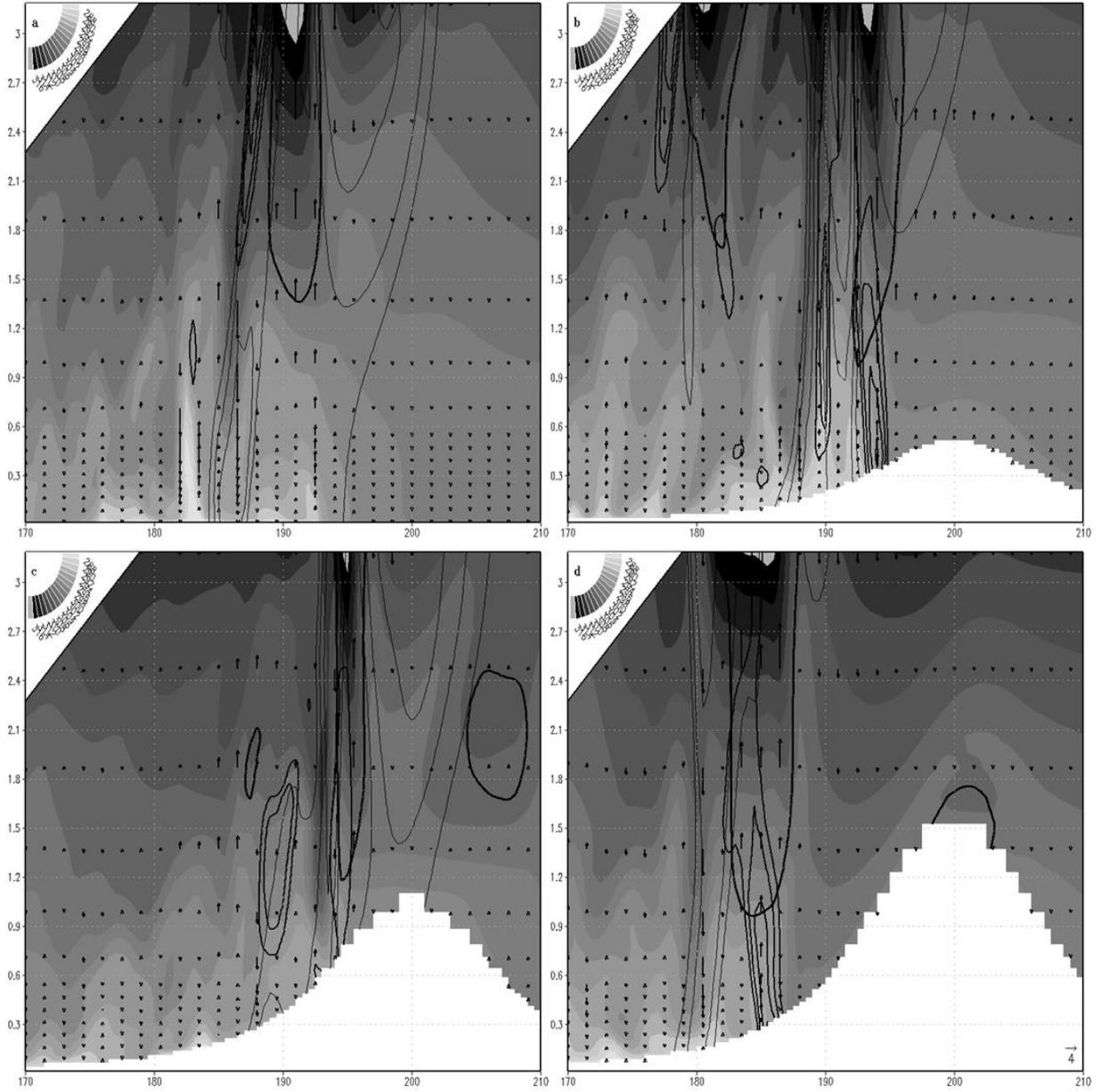

Figure 10: Close up vertical cross section along the east-west mountain ridge at the 180[th] min of simulation time. Theta, Reflectivity (starting at 50 dBZ, thin contours every 5dBZ), Cloud outline (thick contour), and Vertical Vorticity (medium contour, Levels are 0.01, 0.015, 0.02, 0.025 s$^{-1}$)

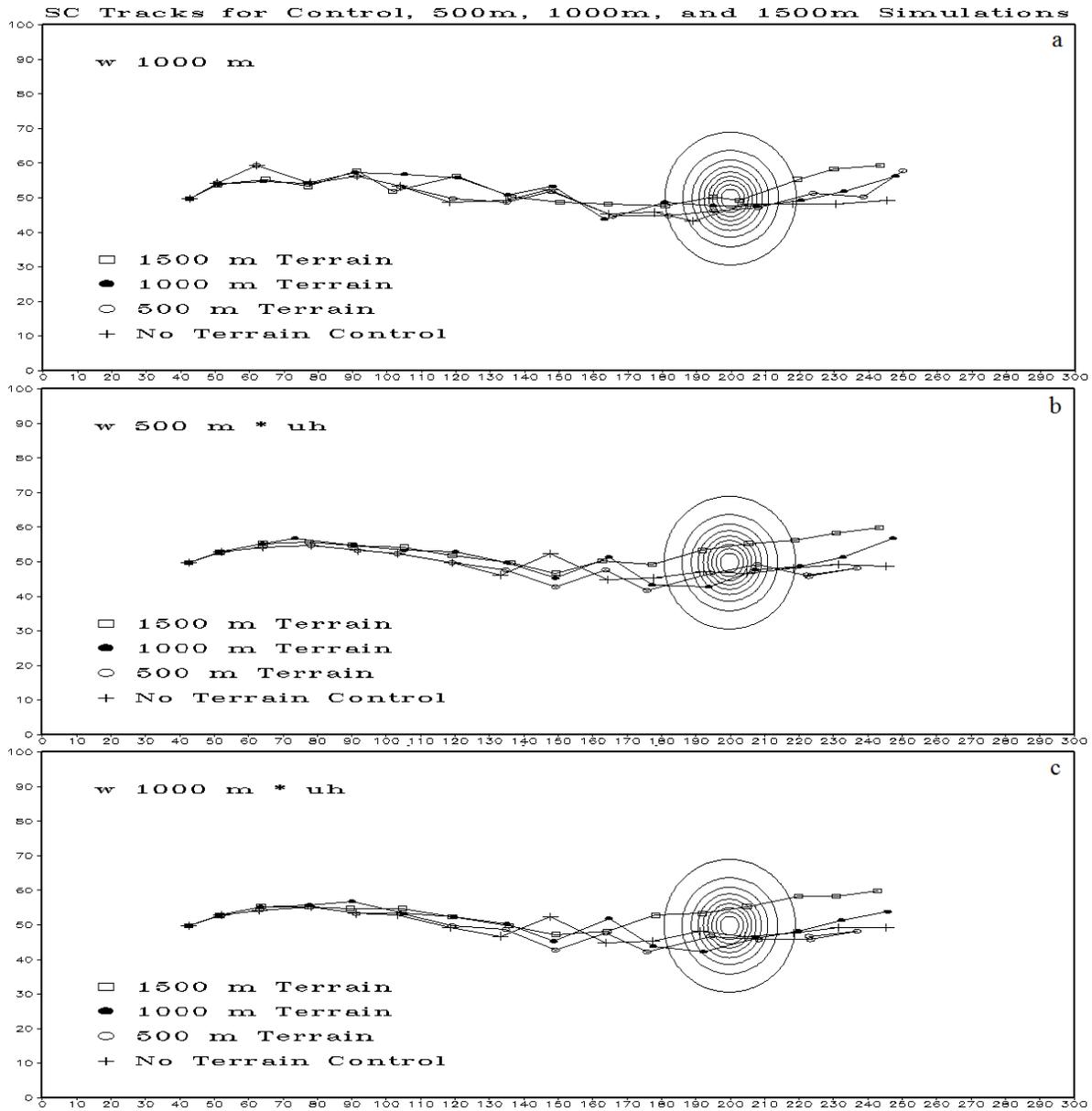

Figure 11: Tracks of supercell thunderstorms as identified by (a) 1000 m AGL Updraft strength (b) Updraft Helicity (UH) multiplied by vertical velocity at 500 m AGL (c) Updraft Helicity multiplied by vertical velocity (UHW) 1000 m AGL. The contours represent the normalized terrain height.